\documentclass[aps,prb,twocolumn,superscriptaddress,showkeys]{revtex4-1}

\usepackage{graphicx}
\usepackage{amsmath}
\begin{document}

\title{EXAFS and electrical studies of new narrow-gap semiconductors: \\
InTe$_{1-x}$Se$_x$ and In$_{1-x}$Ga$_x$Te \\}

\author{A.~I.~Lebedev}
\email[]{swan@scon155.phys.msu.ru}
\author{A.~V.~Michurin}
\author{I.~A.~Sluchinskaya}
\affiliation{Physics Department, Moscow State University, 119899 Russia}
\author{V.~N.~Demin}
\affiliation{Chemistry Department, Moscow State University, 119899 Russia}
\author{I.~H.~Munro}
\affiliation{CCLRC, Daresbury Laboratory, Warrington, WA4 4AD, UK}

\date{\today}

\begin{abstract}
The local environment of Ga, Se, and Tl atoms in InTe-based solid solutions 
was studied by EXAFS technique. It was shown that all investigated atoms
are substitutional impurities, which enter the In(1), Te, and In(2)
positions in the InTe structure, respectively. The electrical measurements
revealed that In$_{1-x}$Ga$_x$Te and InTe$_{1-x}$Se$_x$ solid solutions
become semiconductors at $x>0.24$ and $x>0.15$, respectively.

\texttt{DOI: 10.1016/S0022-3697(00)00196-7}
\end{abstract}


\maketitle

\section{Introduction}

Narrow-gap semiconductors have attracted considerable interest due to their wide
application in infrared optoelectronics. Among these materials, solid solutions
in which the band gap is controlled by the crystal composition have received
particular attention. Our interest to InTe-based solid solutions is stimulated
by a possibility of obtaining new narrow-gap semiconductors as a result of
semimetal--semiconductor transition, which can occur at isoelectronic substitution
in these crystals. In$_{1-x}$Tl$_x$Te is an example of such a solid solution, in
which the substitution of indium by thallium results in appearance of semiconductor
properties at $x > x_c \approx 0.07$.~\cite{ApplPhysLett.38.634,FizTverdTela.28.2680}
In this work we investigate two other solid solutions, In$_{1-x}$Ga$_x$Te and
InTe$_{1-x}$Se$_x$, which crystallize in the same TlSe-type tetragonal structure
with the space group $I4/mcm$ and can be regarded as candidates for narrow-gap
semiconductors.

\begin{figure}
\includegraphics{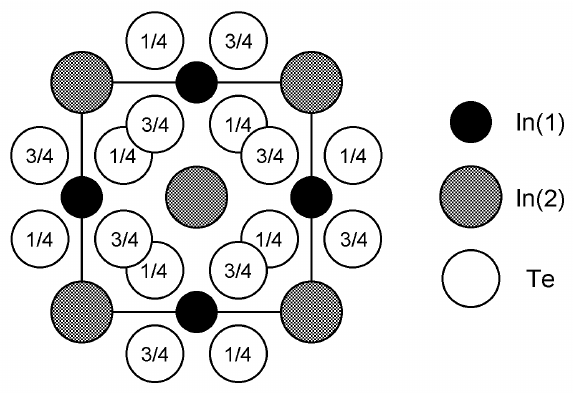}
\caption{\label{fig1} A projection of the InTe structure on the $ab$~plane. The
In atoms form chains going along the $c$ axis and are at the heights of 0 and
1/2 (in units of $c$).}
\end{figure}

The metal atoms in InTe occupy two positions (see Fig.~\ref{fig1}). The In(1)
atoms in position 4($b$) are tetrahedrally coordinated by four chalcogen
atoms and form chains going along the $c$~axis. The bonding in these chains is
predominantly covalent. The In(2) atoms in position 4($a$) also form chains,
but are surrounded by eight chalcogen atoms occupying the position 8($h$) and
two In(2) atoms from the same chain. The metal atoms in the In(1) position are
positively charged (valence +3) and the atoms in the In(2) position are negatively
charged (valence +1).

The purpose of this work was to substitute the atoms in all lattice sites of
InTe and to establish conditions, under which the solid solutions acquire
semiconductor properties. One could suppose that Se atoms would substitute for
those of Te. According to the known crystal structure of InTlTe$_2$%
    \footnote {The InTlTe$_2$ compound is isostructural to InTe and is the
    limiting case of the In$_{1-x}$Tl$_x$Te solid solution, in which all In(2)
    positions are occupied by Tl and all In(1) positions are occupied by In
    (Ref.~\onlinecite{ZAnorgAllgChem.398.207}).}
and the valence +1 typical for the Tl atom, one could expect Tl to substitute
for In in the In(2) position. The Ga atoms, which have typical valence +3 and
small size, should probably enter the In(1) position.

To determine the location of Ga, Se, and Tl atoms in the solid solutions we used
EXAFS technique, which is based on the analysis of the extended fine structure
in X-ray absorption spectra.~\cite{RevModPhys.53.769}  EXAFS studies were
complemented by X-ray diffraction studies, which enabled to establish the
solubility of the impurities in InTe and to determine the variation of the
lattice parameters in the solid solutions.

\section{Samples}

In this work, polycrystalline samples and single crystals of
InTe$_{1-x}$Se$_x$, In$_{1-x}$Ga$_x$Te, and In$_{1-x}$Tl$_x$Te solid
solutions were studied. Polycrystalline samples were prepared by alloying
the appropriate amounts of InTe, InSe, GaTe, and TlTe in evacuated silica
ampoules and were annealed at a temperature by 30$^\circ$C below the solidus
one for 7~days. The lattice parameters $a(x)$ and $c(x)$ for
InTe$_{1-x}$Se$_x$ and In$_{1-x}$Ga$_x$Te solid solutions were measured
on the samples with compositions $x = {}$0, 0.1, 0.2, 0.3, 0.4, 0.5. The
In$_{1/2}$Ga$_{1/2}$Te sample ($x = 0.5$) appeared to be two-phase and
was not taken into consideration. The dependence of $a$ and $c$ on $x$
can be approximated by linear curves,
    \begin{equation}
    a(x) = 8.4343-0.284x,~~c(x) = 7.1452-0.381x, \\
    \label{eq1}
    \end{equation}
    \begin{equation}
    a(x) = 8.4320-0.084x,~~c(x) = 7.1482-0.656x,
    \label{eq2}
    \end{equation}
for InTe$_{1-x}$Se$_x$ and In$_{1-x}$Ga$_x$Te, respectively.

Single crystals of the solid solutions were grown by the Bridgman method.
As the composition of the crystals changed along the ingot, samples with
different $x$ were cut from different parts of the crystal. The
composition of the samples was calculated from the $c$ parameter, which
depended on $x$ stronger than the $a$ parameter. The error in
determination of $x$ did not exceed 0.02. According to the X-ray data,
in InTe$_{1-x}$Se$_x$ crystals the $x$ value increased along the ingot,
while in In$_{1-x}$Ga$_x$Te crystals it slowly decreased.

Before EXAFS measurements the samples were powdered, the powders were
sifted through a sieve and then rubbed into the surface of adhesive
tape. The optimum thickness of the absorbing layer was achieved by
folding the tape (typically 8--16 layers).

Electrical measurements were carried out on the needle-shaped samples
cleaved from the ingots along the $c$ axis. The current and potential
contacts to the samples with characteristic sizes of 1$\times$1$\times$5~mm
were soldered using indium under the NH$_4$Cl flux. The contacts appeared
to be non-ohmic.

\section{Experiment}

\subsection{EXAFS studies}

The method of EXAFS-spectroscopy is based on the study of the extended
fine structure appearing above the X-ray absorption edge and resulting
from the interference of the out-going photoelectron wave exited from
one of the core levels of a central atom under investigation with the waves
scattered by neighboring atoms. In the single-scattering approximation,
the $\chi$ function that describes the dependence of an oscillatory part
of the absorption coefficient on the photoelectron wave vector $k$ is
    \begin{equation}
    \begin{aligned}
    \chi(k) &= - \frac{1}{k} \sum_j \frac{N_j S_0^2}{R_j^2} |f(k)|
    \exp \left (- \frac{2R_j}{\lambda} -2\sigma_j^2 k^2 \right) \\
    &\times \sin(2kR_j + 2\delta_l + \phi_j),
    \label{eq3}
    \end{aligned}
    \end{equation}
where the sum runs over a few nearest shells surrounding the central atom, and
$R_j$, $N_j$, and $\sigma_j^2$ are radius, coordination number, and Debye--Waller
factor for the $j$th shell, respectively.~\cite{RevModPhys.53.769}
The parameter $S_0^2$ describes the reduction of the oscillation amplitude
resulting from multi-electron and inelastic scattering effects. The backscattering
amplitude $f(k)$ and phase shift $\phi(k)$, the phase shift of the central
atom $\delta_l(k)$, and the mean free path of a photoelectron $\lambda(k)$ can
be calculated theoretically. By minimizing the deviation between the
theoretical and experimental EXAFS curves, it is possible to find parameters
$R_j$, $N_j$, and $\sigma_j^2$, which describe the position of atoms in several
nearest shells about the investigated atom.

EXAFS measurements were carried out on station 7.1 of the Daresbury synchrotron
radiation source operating at an electron beam energy of 2~GeV and a maximum
stored current of 240~mA. EXAFS data were collected at the $K$~edges of Ga
(10.367~keV) and Se (12.658~keV), and at the $L_{\rm III}$~edge of Tl
(12.658~keV) in transmission mode using ion chambers. The sample temperature
was 80~K. A double Si(111) crystal monochromator was used without the
harmonics rejection, which was not necessary in the above energy range.
For each sample at least two spectra were taken.

The function $\chi(k)$ was extracted from the absorption curve $\mu x(E)$ in
the normal way.~\cite{RevModPhys.53.769}  After removing the pre-edge background,
splines were used to extract the smooth atomic part of the absorption,
$\mu x_{0}(E)$, and the dependence $\chi=(\mu x-\mu x_{0})/\mu x_{0}$
was calculated as a function of the photoelectron wave vector
$k=[2m(E-E_{0})/\hbar^2]^{1/2}$. The energy origin, $E_{0}$, was taken to be
at the inflection point on the absorption edge. The edge steps ranged from
0.06 to 0.41.

\begin{table*}
\caption{\label{table1}Parameters of the local structure for InTe-based solid solutions.}
\begin{ruledtabular}
\begin{tabular}{llllllll}
Sample & InTe\footnotemark[1] & In$_{0.8}$Ga$_{0.2}$Te & InTe$_{0.9}$Se$_{0.1}$ & InTe$_{0.8}$Se$_{0.2}$ & InTe$_{0.7}$Se$_{0.3}$ & In$_{0.8}$Tl$_{0.2}$Te\footnotemark[2] & InTlTe$_2$\footnotemark[2] \\
\hline
Standard & InSb & Ga$_2$Te$_3$ & InSe & InSe & InSe & TlI & TlI \\
\hline
$R_1$,~{\AA} & 2.821(2) & 2.643(4) & 2.608(4) & 2.607(3) & 2.605(3) & 3.525(12) & 3.564(11) \\
$N_1$ & 2.1(1) & 4.2(3) & 1.7(2) & 1.8(1) & 1.7(1) & 10.2(15) & 13.0(18) \\
$\sigma_1^2$,~{\AA$^2$} & 0.0021(3) & 0.0043(5) & 0.0009(5) & 0.0016(4) & 0.0013(4) & 0.0124(19) & 0.0143(16) \\
\hline
$R_2$,~{\AA} & 3.541(8) & 3.497(28) & 3.557(35) & 3.526(43) & 3.528(53) & 4.271(34) & 4.262(23) \\
$N_2$ & 6.4(4) & 2.1(2) & 3.4(3) & 3.5(2) & 3.3(2) & 4.1(6) & 5.2(7) \\
$\sigma_2^2$,~{\AA$^2$} & 0.0117(10) & 0.0038(40) & 0.030(7) & 0.049(10) & 0.055(13) & 0.0180(68) & 0.0168(44) \\
\hline
$R_3$,~{\AA} & 4.215(12) & 4.272(28) & 3.834(17) & 3.823(15) & 3.821(16) & 5.664(32) & 5.698(30) \\
$N_3$ & 6.4(4) & 4.2(3) & 1.7(2) & 1.8(1) & 1.7(1) & 8.2(12) & 10.4(14) \\
$\sigma_3^2$,~{\AA$^2$} & 0.0136(19) & 0.0179(56) & 0.0031(17) & 0.0027(14) & 0.0015(15) & 0.0139(52) & 0.0165(48) \\
\end{tabular}
\end{ruledtabular}
\footnotetext[1]{EXAFS data at the In $K$~edge for InTe were collected on station
X23A2 at NSLS. The presented coordination numbers correspond to averaged number
of neighbors per In atom.}
\footnotetext[2]{It was assumed that eight Te and two metal atoms in the first
shell of thallium are at the same distance and have equal Debye--Waller factors.}
\end{table*}

\begin{figure*}
\includegraphics{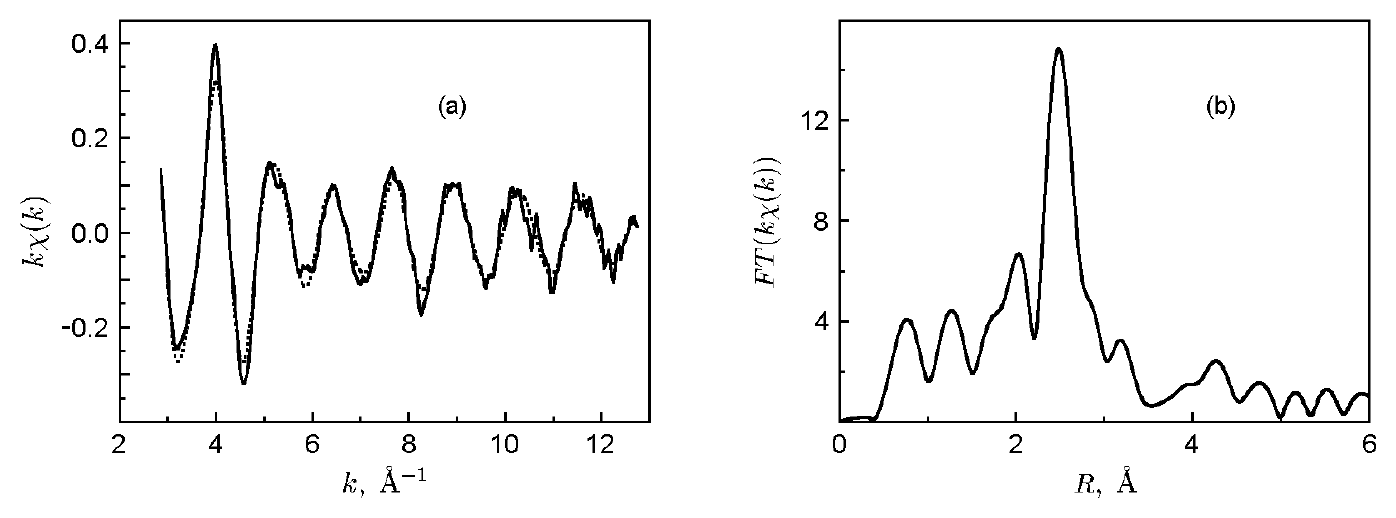}
\caption{\label{fig2}(a) Typical EXAFS spectrum obtained at the Ga $K$~edge for
In$_{0.8}$Ga$_{0.2}$Te sample. Experimental data are shown by the solid line
while the dashed line is their best theoretical approximation. (b) The magnitude
of the Fourier transform of the $k \chi(k)$ data. Three first shells were
isolated by inverse Fourier transform in the $1.5 \le R \le 4.55$~{\AA} range.}
\end{figure*}

\begin{figure*}
\includegraphics{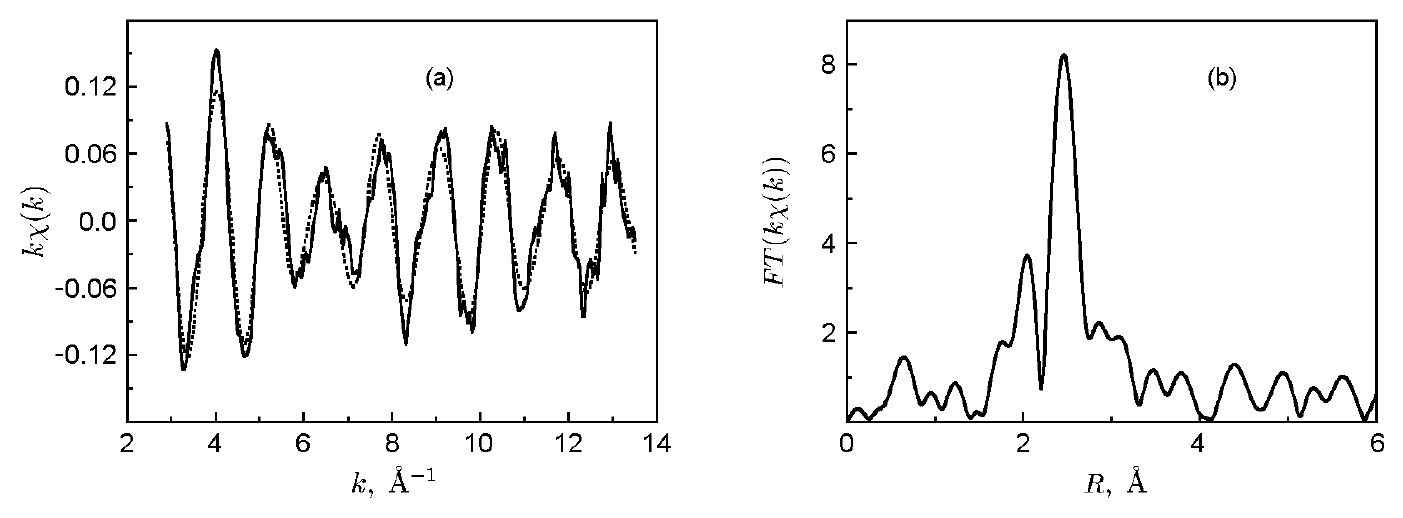}
\caption{\label{fig3}(a) Typical EXAFS spectrum obtained at the Se $K$~edge for
InTe$_{0.8}$Se$_{0.2}$ sample. Experimental data are shown by the solid line
while the dashed line is their best theoretical approximation. (b) The magnitude
of the Fourier transform of the $k \chi(k)$ data. Three first shells were
isolated by inverse Fourier transform in the $1.55 \le R \le 4.1$~{\AA} range.}
\end{figure*}

\begin{figure*}
\includegraphics{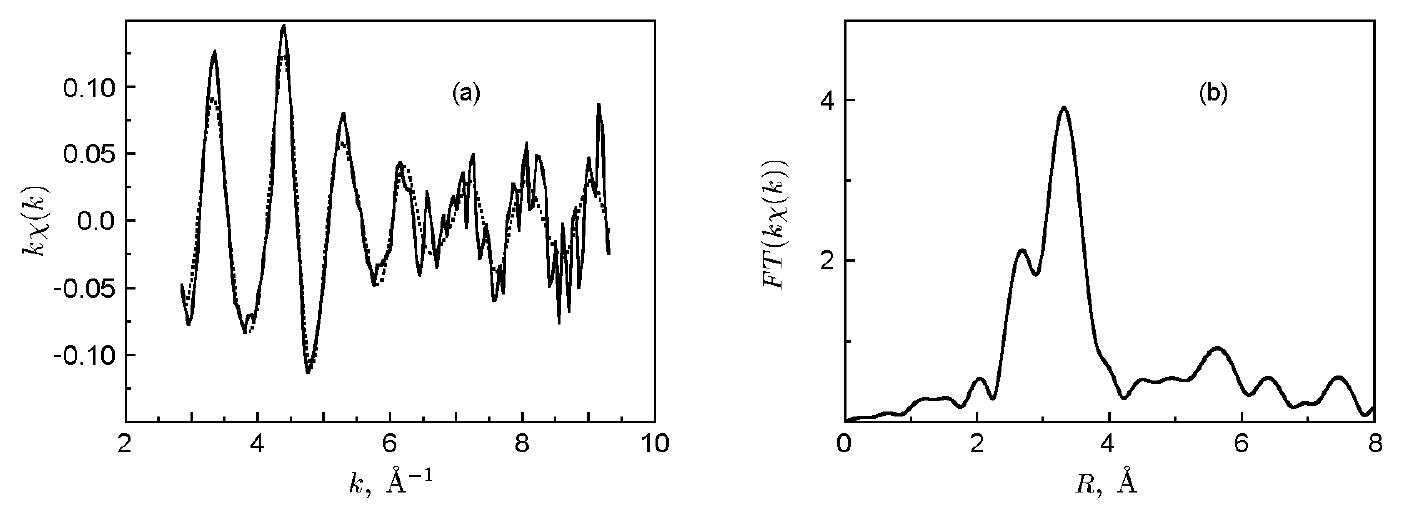}
\caption{\label{fig4}(a) Typical EXAFS spectrum obtained at the Tl $L_{\rm III}$~edge
for In$_{0.8}$Tl$_{0.2}$Te sample. Experimental data are shown by the
solid line while the dashed line is their best theoretical approximation.
(b) The magnitude of the Fourier transform of the $k \chi(k)$ data. Three
first shells were isolated by inverse Fourier transform in the
$2.25 \le R \le 6.2$~{\AA} range.}
\end{figure*}

Direct and inverse Fourier transforms with modified Hanning windows were used to
extract the information about the first three shells from the experimental curve
$\chi(k)$. The ranges of extraction in $R$~space are presented in the captions
for Figs. 2--4. The parameters $R_{j}$, $N_{j}$, and $\sigma_{j}^{2}$ for
each shell as well as the energy origin correction $\delta E_{0}$ were
simultaneously varied to obtain the minimum root-mean-square deviation between
the experimental and calculated $k\chi(k)$ curves. The \texttt{FEFF}
software~\cite{PhysRevB.44.4146} was used to calculate the $f(k)$, $\phi(k)$,
$\delta_l(k)$, and $\lambda(k)$ functions needed for Eq.~(\ref{eq3}). The
number of the fitting parameters (8) was usually less than a half of the number
of independent data points $N_{ind}=2\Delta k\Delta R/\pi={}$16--21. The accuracy
of determination of fitting parameters was estimated from the correlation matrix;
the errors presented in Table~\ref{table1} are the 95\% confidence intervals for
variation in the parameters resulting from statistical errors in experimental data.
To increase the accuracy, the energy corrections $\delta E_{0}$ for all shells was
assumed to be the same, and known relation between coordination numbers for the
InTe structure was used.

For each absorption edge the $S_0^2$ values, which are necessary for determination
of coordination numbers, were obtained from the studies of reference compounds
(InSe for the Se edge, Ga$_2$Te$_3$ for the Ga edge, and TlI for the Tl edge).

A typical $k \chi(k)$ curve obtained at the Ga $K$~edge for
In$_{0.8}$Ga$_{0.2}$Te sample is shown in Fig.~\ref{fig2}(a), and its Fourier
transform is shown in Fig.~\ref{fig2}(b). An analysis of the data reveals that
in the first shell the Ga atom is surrounded by four Te atoms located
at a distance of 2.64~{\AA} (Table~\ref{table1}). It indicates that
the Ga atoms occupy the In(1) position in the solid solution. The obtained
Ga--Te distance is close to that found by us in Ga$_2$Te$_3$ (2.61~{\AA}),
in which Ga is tetrahedrally coordinated by tellurium; this can indicate
the covalent character of the Ga--Te bond in the solid solution. The position
of Ga atoms is confirmed by the observation of two metal atoms from the same
chain (shifted along the $c$~axis) and four metal atoms lying in the
perpendicular direction (the second and third shells, see Table~\ref{table1}).
The distances to these atoms, determined from EXAFS, agree well with the
X-ray diffraction data; they are slightly shorter than the In--In distances
in InTe. The Ga--Te bond is 0.18~{\AA} shorter than the In(1)--Te bond
in InTe.

A typical EXAFS spectrum obtained at the Se $K$~edge for InTe$_{0.8}$Se$_{0.2}$
sample and its Fourier transform are presented in Fig.~\ref{fig3}. An analysis
of these data reveals that the Se atoms are surrounded by two In atoms at
2.61~{\AA} (Table~\ref{table1}). The length of this bond is close to that of the
In--Se covalent bond in InSe (2.615~{\AA}); this enables to conclude that the
In--Se bond in the solid solution is also covalent. The obtained coordination
number and the bond length indicate that Se atoms substitute for Te in the
solid solution. The Se--In(1) distance is 0.22~{\AA} shorter than the Te--In(1)
distance in InTe. The second shell of the Se atom (four atoms in the In(2)
position) is strongly distorted and is characterized by an unexpectedly large
Debye--Waller factor, which increases with increasing $x$ (Table~\ref{table1}).
However, the analysis reveals an appreciable contribution from the atoms in the
third shell (two closest chalcogen atoms in a distorted square antiprism).

A typical EXAFS spectrum obtained at the Tl $L_{\rm III}$~edge for
In$_{0.8}$Tl$_{0.2}$Te sample and its Fourier transform are shown in
Fig.~\ref{fig4}. An analysis reveals that the Tl atoms substitute for In
in the In(2) position. It is interesting that the Tl--Te distances in
In$_{0.8}$Tl$_{0.2}$Te and InTlTe$_2$ are very close to the In(2)--Te
distance in InTe (Table~\ref{table1}). Another feature of these crystals
is a rather high Debye--Waller factor for the Tl--Te bond in the first shell.

\subsection{Electrical properties}

To study possible band structure changes we have measured the temperature
dependence of specific resistivity $\rho(T)$ on InTe$_{1-x}$Se$_x$ and
In$_{1-x}$Ga$_x$Te solid solutions.%
    \footnote{Electrical properties of the
    In$_{1-x}$Tl$_x$Te solid solution were studied in detail in
    Refs.~\onlinecite{ApplPhysLett.38.634,FizTverdTela.28.2680,FizTverdTela.29.16}.}
According to the sign of the Seebeck coefficient, all the samples were $p$-type.

\begin{figure}
\includegraphics{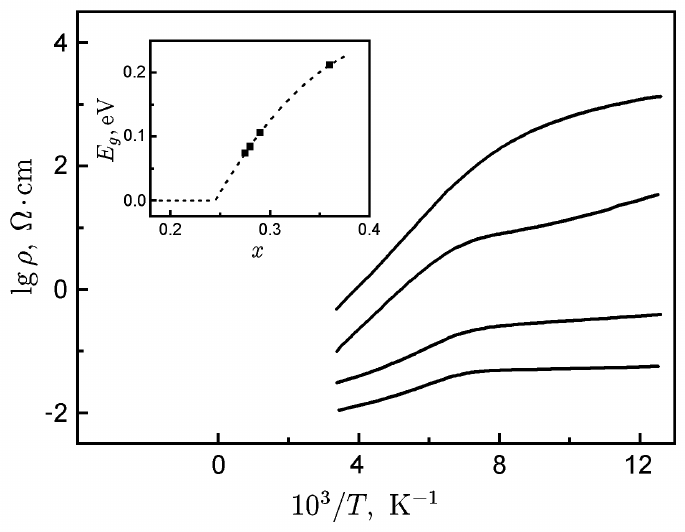}
\caption{\label{fig5}Temperature dependence of specific resistivity for
In$_{1-x}$Ga$_x$Te samples. The insert shows the dependence of the band
gap on the composition of the solid solution.}
\end{figure}

Typical $\log_{10}\rho(10^3/T)$ plots for In$_{1-x}$Ga$_x$Te samples are shown
in Fig.~\ref{fig5}. For all investigated samples, the $\rho(T)$ curves had an
activation (semiconductor) character. Two regions with different slopes and
a break at about 140~K are clearly seen on the curves. We associated the
high-temperature region with intrinsic conductivity, and the low-temperature one
with the impurity or hopping conductivity. The forbidden energy gap $E_g(x)$,
calculated as the doubled activation energy in the high-temperature region, is
plotted as a function of $x$ in the insert to Fig.~\ref{fig5}. The extrapolation
of this dependence to $E_g = 0$ gives the composition $x_c \approx 0.24$, above
which the In$_{1-x}$Ga$_x$Te crystals become semiconductors.

\begin{figure}
\includegraphics{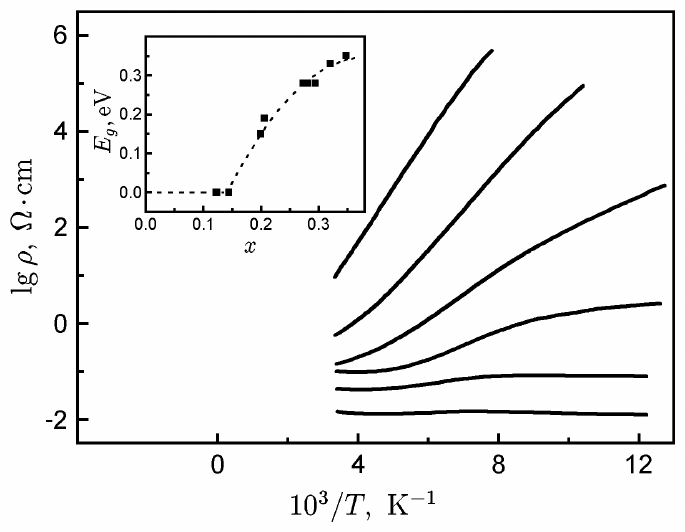}
\caption{\label{fig6}Temperature dependence of specific resistivity for
InTe$_{1-x}$Se$_x$ samples. The insert shows the dependence of the band
gap on the composition of the solid solution.}
\end{figure}

Typical $\log_{10}\rho(10^3/T)$ plots for InTe$_{1-x}$Se$_x$ samples are
presented in Fig.~\ref{fig6}. Among these crystals there were the samples, for
which the $\rho(T)$ curves had the semiconductor or the metallic character.
The energy gap $E_g(x)$, calculated as the doubled activation energy in the
region with the greatest slope, is plotted as a function of $x$ in the insert
to Fig.~\ref{fig6}. It is seen that the InTe$_{1-x}$Se$_x$ crystals are
semiconductors at $x > x_c \approx 0.15$.

\section{Discussion}

We use the obtained EXAFS and X-ray diffraction data to estimate the position
of all atoms in the InTe structure.

As follows from the EXAFS data, the Ga and Tl atoms occupy, respectively,
highly symmetric In(1) and In(2) positions in InTe.

The position of the chalcogen atom in the structure is described by the
position parameter~$u$. Unfortunately, in solid solutions the determination
of this parameter is impossible because of the random character of atomic
substitution. That is why we considered the end-point compounds, on the
base of which the solid solutions are formed. These were: the hypothetical
In$_{1/2}$Ga$_{1/2}$Te compound,%
    \footnote{An attempt to determine the structure of this compound was
    undertaken in Ref.~\onlinecite{ZAnorgAllgChem.525.163}. However, the
    authors themselves noted that their samples were not single-phase.}
in which all In(1) positions are occupied by Ga, and the hypothetical
InSe compound with the InTe structure. To determine~$u$, we used the
lattice parameters and the shortest interatomic distances (Ga--Te or
In(1)--Se) obtained from the EXAFS measurements. The lattice parameters
for hypothetical compounds were estimated from Eqs.~(\ref{eq1}) and
(\ref{eq2}). The shortest interatomic distances in the compounds
were assumed to be the same as in solid solutions. This assumption is
based on the well-known fact of a weak dependence of the chemical bond
length on the composition of solid solution, which is partly confirmed
by our experimental data.

It was found that in In$_{1/2}$Ga$_{1/2}$Te the position parameter
of the Te atom is $u_{\rm Te} = 0.170$, and the calculated In(2)--Te
distance is 3.55~{\AA}, which remains nearly the same as in InTe.
In hypothetical InSe compound with the InTe structure, the parameter
$u_{\rm Se}$ is 0.172, and the calculated In(2)--Se distance is
3.46~{\AA}. It should be noted that the substitution of Te by Se results
in contraction of the In(1)--chalcogen distance by 0.22~{\AA}, whereas the
In(2)--chalcogen distance shortens by only $\approx$\,0.1~{\AA}.

Our EXAFS data for InTlTe$_2$ agree well with the X-ray structure
data (Ref.~\onlinecite{ZAnorgAllgChem.398.207}).%
    \footnote{Small difference between the Tl--Te distances obtained from the
    EXAFS data (3.56~{\AA}) and X-ray data (3.595~{\AA}) can be explained,
    as in Ref.~\onlinecite{PhysRevB.55.14770}, by a systematic error in
    calculation of the phase shifts in the \texttt{FEFF} program for heavy
    atoms. For Pb, which is adjacent to Tl in the Periodic Table, the EXAFS
    distances were by 0.025~{\AA} shorter than the X-ray
    ones (Ref.~\onlinecite{PhysRevB.55.14770}).}
As was mentioned above, in In$_{1-x}$Tl$_x$Te the distance between the
metal in the In(2) position and Te atoms changes a little when substituting
In atoms by Tl. In any case, the observed change is notably less than
the difference in ionic radii of Tl$^+$ and In$^+$ (0.12~{\AA} according
to Ref.~\onlinecite{LangeHandbookChemistry.1973}).

The obtained parameters $u$ for the chalcogen atom enable to estimate
the distortions of structure of the solid solutions. It is known that
the tetrahedron that surrounds the In(1) atom in InTe is slightly
stretched along the $c$~axis, and the Te--In(1)--Te angle is
$\theta = {}$101$^\circ$\,15$'$. Our calculations show that in all
studied solid solutions the distortion of the tetrahedron increases
compared to that in InTe: it is minimal in In$_{1/2}$Tl$_{1/2}$Te
($\theta = {}$101$^\circ$\,1$'$), more appreciable in In$_{1/2}$Ga$_{1/2}$Te
($\theta = {}$99$^\circ$\,31$'$), and is maximal in hypothetical InSe with
the InTe structure ($\theta = {}$99$^\circ$\,2$'$).

Our results enable to understand the role of the In(2) position
in the InTe structure. Large Debye--Waller factors for the In(2)--Te
distance in InTe and InTe$_{1-x}$Se$_x$ and for the Tl--Te distance in
In$_{1-x}$Tl$_x$Te (Table~\ref{table1}), unexpectedly small change of the
distance between the metal atoms in the In(2) position and the chalcogen
atoms when substituting In\,$\to$\,Tl ($<$0.04~{\AA}) and Te\,$\to$\,Se (about
0.1~{\AA}) are evidences of a weak chemical bond between the metal atoms
in this position and neighboring atoms. This can indicate that the In(2)
position in the structure plays a role of a ``cavity'' for single-charged
ions, necessary to maintain the electrical neutrality of a sample. This
explains why the InTe structure remains unchanged when substituting the
In(2) atoms by various single-charged ions with strongly different radii
(Na, K, Tl).

Strong increase in the Debye--Waller factor for the In(2)--chalcogen
distance in InTe$_{1-x}$Se$_x$ is apparently associated with appreciable
displacement of the In(2) atoms from highly symmetric position as a
result of a reduced symmetry of their environment (random arrangement
of the chalcogen atoms). As follows from smaller Debye--Waller factors for
the In(1)--Se and Se--chalcogen distances, the displacements of the In(1)
and chalcogen atoms in the solid solution are not so large.

A comparison of the electrical properties of different InTe-based solid solutions
revealed a peculiarity in the behavior of $\rho(T)$ curves in the semiconductor
samples of InTe$_{1-x}$Se$_x$: in the high-temperature region they appreciably
deviated from the expected activation dependence. Such a deviation was not
observed in In$_{1-x}$Tl$_x$Te (Ref.~\onlinecite{FizTverdTela.28.2680}) and
In$_{1-x}$Ga$_x$Te. We suppose that this deviation is due to a combined effect
of a weak chemical bonding characteristic for the atoms in the In(2) position and
of increased In(2)--Se distance in the hypothetical InSe compound. They can lead
to an appreciable increase in the size of a cavity for In(2) in InTe$_{1-x}$Se$_x$.
As a consequence, the anharmonicity of the motion of the In(2) atoms increases
and the temperature variation of the band gap becomes non-linear, thus resulting
in the deviation of $\rho(T)$ curves from the activation dependence.

In conclusion, we would like to note that the appearance of semiconductor
properties as a result of atomic substitution in all lattice sites is a common
feature of all studied InTe-based solid solutions. We can offer the following
qualitative explanation of this fact. The character of filling of the energy
bands in InTe is close to that in semiconductors, but owing to individual
features of its atomic components, the bottom of the conduction band appears
slightly below the top of the valence band, thus forming a semimetal band
structure. In the case of In$_{1-x}$Ga$_x$Te and InTe$_{1-x}$Se$_x$, the
strengthening of the covalent bond (as a consequence of reduction of interatomic
distances in tetrahedra) increases the splitting of hybridized orbitals, thus
causing the appearance of a gap in the electronic spectrum and the onset
of semiconductor behavior. In the case of In$_{1-x}$Tl$_x$Te, the covalent
bond length remains unchanged, and the cause of the appearance of semiconductor
properties is different. As the contribution of $s$ states to the valence band
is usually higher than to the conduction band, and the Tl $6s$ states lie deeper
than the In $5s$ states due to strong relativistic corrections, the valence
band moves downwards when substituting In\,$\to$\,Tl. As a result, the energy
gap arises in the electronic spectrum, and the solid solution becomes a
semiconductor.


\providecommand{\BIBYu}{Yu}

\end{document}